\documentclass[aps,prl,superscriptaddress,twocolumn,showpacs,preprintnumbers,amsmath,amssymb]{revtex4-2}

\usepackage{graphicx}
\usepackage{dcolumn}
\usepackage{bm}
\usepackage{color}
\usepackage[dvipsnames]{xcolor}

\begin{document}

\preprint{}

\title{Skyrmion-(anti)vortex coupling in a chiral magnet-superconductor heterostructure}

\author{A.P. Petrovi\'c}
\email{appetrovic@ntu.edu.sg}
\affiliation{Division of Physics and Applied Physics, School of Physical and Mathematical Sciences, Nanyang Technological University, 637371 Singapore.}
\author{M. Raju}
\altaffiliation[Present address: ]{Institute for Quantum Matter and Department of Physics and Astronomy, Johns Hopkins University, Baltimore, MD 21218, USA.}
\affiliation{Division of Physics and Applied Physics, School of Physical and Mathematical Sciences, Nanyang Technological University, 637371 Singapore.}
\author{X.Y. Tee}
\affiliation{Division of Physics and Applied Physics, School of Physical and Mathematical Sciences, Nanyang Technological University, 637371 Singapore.}
\author{A. Louat}
\affiliation{Department of Physics, Technion, Haifa 32000, Israel.}
\author{I. Maggio-Aprile}
\affiliation{Department of Quantum Matter Physics, Universit\'{e} de Gen\`{e}ve, 24 Quai Ernest Ansermet, CH-1211 Geneva 4, Switzerland.}
\author{R.M. Menezes}
\affiliation{Department of Physics, University of Antwerp, Groenenborgerlaan 171, B-2020 Antwerp, Belgium.}
\affiliation{Departamento de F\'{\i}sica, Universidade Federal de Pernambuco, Cidade Universit\'{a}ria, 50670-901, Recife-PE, Brazil.}
\author{M.J.~Wyszy\'{n}ski}
\affiliation{Department of Physics, University of Antwerp, Groenenborgerlaan 171, B-2020 Antwerp, Belgium.}
\author{N.K. Duong}
\affiliation{Division of Physics and Applied Physics, School of Physical and Mathematical Sciences, Nanyang Technological University, 637371 Singapore.}
\author{M. Reznikov}
\affiliation{Department of Physics, Technion, Haifa 32000, Israel.}
\author{Ch. Renner}
\affiliation{Department of Quantum Matter Physics, Universit\'{e} de Gen\`{e}ve, 24 Quai Ernest Ansermet, CH-1211 Geneva 4, Switzerland.}
\author{M.V. Milo\v{s}evi\'c}
\affiliation{Department of Physics, University of Antwerp, Groenenborgerlaan 171, B-2020 Antwerp, Belgium.}
\author{C. Panagopoulos}
\email{christos@ntu.edu.sg} 
\affiliation{Division of Physics and Applied Physics, School of Physical and Mathematical Sciences, Nanyang Technological University, 637371 Singapore.}

\date{\today}

\begin{abstract}
We report experimental coupling of chiral magnetism and superconductivity in [IrFeCoPt]/Nb heterostructures. The stray field of skyrmions with radius $\approx50$\,nm is sufficient to nucleate antivortices in a 25\,nm Nb film, with unique signatures in the magnetization, critical current and flux dynamics, corroborated via simulations. We also detect a thermally-tunable Rashba-Edelstein exchange coupling in the isolated skyrmion phase. This realization of a strongly interacting skyrmion-(anti)vortex system opens a path towards controllable topological hybrid materials, unattainable to date.
\end{abstract}


\maketitle

\textit{Introduction.}\textemdash~Chiral magnets and superconductors host topological excitations known as skyrmions and vortices, whose duality was recognised in the 1980s~\cite{Bogdanov1989}. Recently, skyrmion-vortex pairing in heterostructures~\cite{Hals2016,Baumard2019,Dahir2019,Menezes2019,Garnier2019a} has been proposed as a method of coupling chiral magnetism and superconductivity: a combination anticipated to deliver novel hybrid behavior~\cite{Yokoyama2015,Pershoguba2016}. For example, one can envisage changing the superconducting order parameter symmetry by imprinting non-collinear exchange fields from skyrmions or spin helices onto Cooper pairs. Such fields are gauge-equivalent to a Zeeman field combined with spin-orbit coupling (SOC) and may hence create a topological superconductor hosting Majorana fermions at its boundaries and vortex cores~\cite{Nakosai2013,Chen2015c,Yang2016b,Gungordu2018,Rex2019,Garnier2019,Zhou2019,Rex2020}. Controlling the nucleation and dynamics of vortices in the presence of skyrmions is the key to unlocking the potential of chiral magnet-superconductor hybrids for topological quantum computation~\cite{Posske2020,Ma2020} and fluxonics~\cite{Wang2018h}.

Skyrmions and vortices interact via two mechanisms. The first originates from the Rashba-Edelstein effect~\cite{Edelstein1995}: the skyrmion exchange field combines with interfacial SOC to induce circulating spin-polarized supercurrents, which interfere with vortex currents~\cite{Hals2016,Hals2017,Takashima2016,Hals2016a,Baumard2019}. This interaction (``exchange coupling'') requires contact between superconductor and magnet, and depends on the sign and magnitude of the SOC and exchange field. The second mechanism - stray field coupling - allows skyrmion-vortex interaction without electronic contact at distances greater than the exchange length~\cite{Dahir2019}. The sign and magnitude of this interaction are determined by the current profile induced in the superconductor, which depends on the magnetic layer thickness $d_m$, skyrmion chirality and skyrmion-vortex separation~\cite{Dahir2020}. The skyrmion core polarization is antiparallel to the applied magnetic field $H$ and hence repels vortices. However, a sufficiently large skyrmion can nucleate an antivortex in a nearby superconductor, creating a bound pair of topological solitons experimentally unexplored to date.

\begin{figure*}[htbp]
\includegraphics[clip=true, width=2\columnwidth]{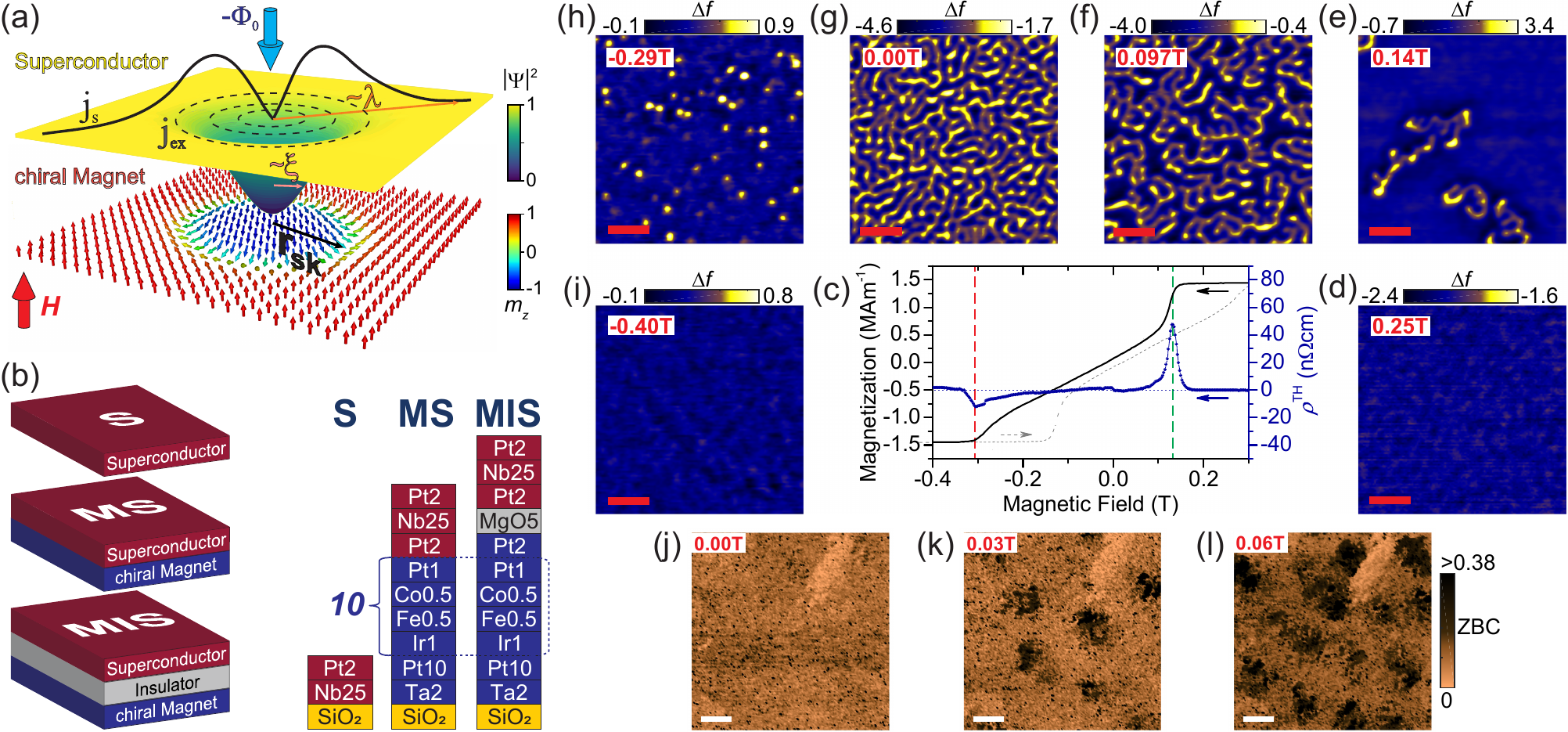}
\caption{\label{Fig1}(a) Schematic of a N{\'e}el skyrmion creating an antivortex with flux $-\Phi_0\equiv-h/2e$ antiparallel to the external magnetic field $H$. Antivortex currents $j_s$ flow at radii up to $\lambda$; Rashba-Edelstein currents $j_{ex}$ are maximal at $r_{sk}$, where the local out-of-plane magnetic moment $m_z=0$. The superconducting order parameter $\left|\Psi\right|$ is suppressed over a length $\xi$ in the vortex core. (b) Sample compositions: numbers (e.g. Ir1, Pt2) indicate layer thicknesses in nm and there are 10 stacked repeats of the [Ir$_{1}$Fe$_{0.5}$Co$_{0.5}$Pt$_1$] unit. (c) Magnetization $M(H)$ and topological Hall resistivity $\rho^{TH}(H)$ at temperature $T=5$\,K for a [Ir$_1$Fe$_{0.5}$Co$_{0.5}$Pt$_1$]$^{10}$ film. Arrows indicate field sweep directions; green/red dashed lines indicate skyrmion nucleation/annihilation fields $H_\mathrm{nuc}$/$H_\mathrm{ann}$ respectively, and the saturation magnetization $M_s=1.45$\,MAm$^{-1}$. (d-i) MFM images at $T = 5.5$~K in a bare [Ir$_1$Fe$_{0.5}$Co$_{0.5}$Pt$_1$]$^{10}$ film during a $H=0.25$\,T~$\rightarrow-0.4$\,T sweep. Scalebars are 500\,nm; colorbars indicate the MFM probe resonance shift $\Delta f$ in Hz, proportional to $m_z$. (j-l) STS images of a 25\,nm Nb film (S-type structure) in (j) $H=$~0\,mT, (k) 30\,mT, and (l) 60\,mT at $T=$~0.4\,K. Scalebars are 100\,nm; the colorbar shows the normalized zero bias conductance (see SM IV) which rises inside the vortex cores.}
\end{figure*}

These interactions can be modulated by adjusting the skyrmion/vortex lengthscales. Stray field coupling is enhanced by increasing skyrmion size, whereas exchange coupling requires Rashba-Edelstein and (anti)vortex currents to circulate with similar radii. This corresponds to the condition $\xi < r_{sk} < \lambda$ (Fig.~\ref{Fig1}(a),~\cite{Baumard2019}), where $\xi$, $\lambda$ are the superconducting coherence and penetration lengths, and $r_{sk}$ is the skyrmion radius. Minimizing $H$ (and hence the vortex density) also favours antivortex formation~\textendash~but aside from Co/Ru(0001) monolayers~\cite{Herve2018}, stabilizing skyrmions at low temperature in thin films has typically required $H\gtrsim1$\,T ~\cite{Romming2015,Kubetzka2020}. This exceeds the upper critical field $H_{c2}$ of many $s$-wave superconductors, precluding skyrmion-(anti)vortex coupling. Here we present the first chiral magnet-superconductor heterostructures to host stable skyrmions at low fields \emph{and} temperatures below the superconducting transition $T_c$. Experiments and simulations both indicate that skyrmion stray fields create antivortices in the superconductor, strongly coupling spin and flux topologies. We also detect signatures of skyrmion-antivortex exchange coupling and identify routes to optimize this effect.  

\textit{Heterostructure design.}\textemdash~MS (chiral Magnet-Superconductor) samples comprise a Nb layer deposited directly onto a [Ir$_{1}$Fe$_{0.5}$Co$_{0.5}$Pt$_1$]$^{10}$ film [schematized in Fig.~\ref{Fig1}(b)], whereas MIS (chiral Magnet-Insulator-Superconductor) samples include a 5\,nm insulating MgO layer to suppress exchange coupling. Fabrication details are in section 1 of the Supplemental Material (SM I).  

\textit{Magnetic layer.}\textemdash~Our selection of [Ir$_{1}$Fe$_{0.5}$Co$_{0.5}$Pt$_1$]$^{10}$ is explained in SM II\,A and its magnetic properties are summarized in Fig.~\ref{Fig1}(c). Upon reducing the field from ferromagnetic saturation, a drop in the magnetization $M(H)$ coincides with a rise in the topological Hall resistivity $\rho^{TH}(H)$, signifying skyrmion nucleation at $H_\mathrm{nuc}$. Magnetic force microscopy (MFM) images acquired in downward field sweeps [Fig.~\ref{Fig1}(d-i)] reveal isolated skyrmions at fields close to the peaks in $\rho^{TH}(H)$ [Fig.~\ref{Fig1}(e)]. At lower fields, the skyrmions proliferate and coalesce into worm-like structures [Fig.~\ref{Fig1}(f)], merging into labyrinthine stripes close to zero field [Fig.~\ref{Fig1}(g)]. After field reversal, stripes eventually split into individual skyrmions [Fig.~\ref{Fig1}(h)] with a concomitant peak in $\rho^{TH}(H)$, prior to erasure at $H_\mathrm{ann}$ in agreement with previous studies~\cite{Soumyanarayanan2017,Raju2019,Duong2019}. MFM image analysis (SM II\,B) indicates a typical $r_{sk}=51\pm6$\,nm.  

\begin{figure}[b]
\includegraphics[clip=true, width=1\columnwidth]{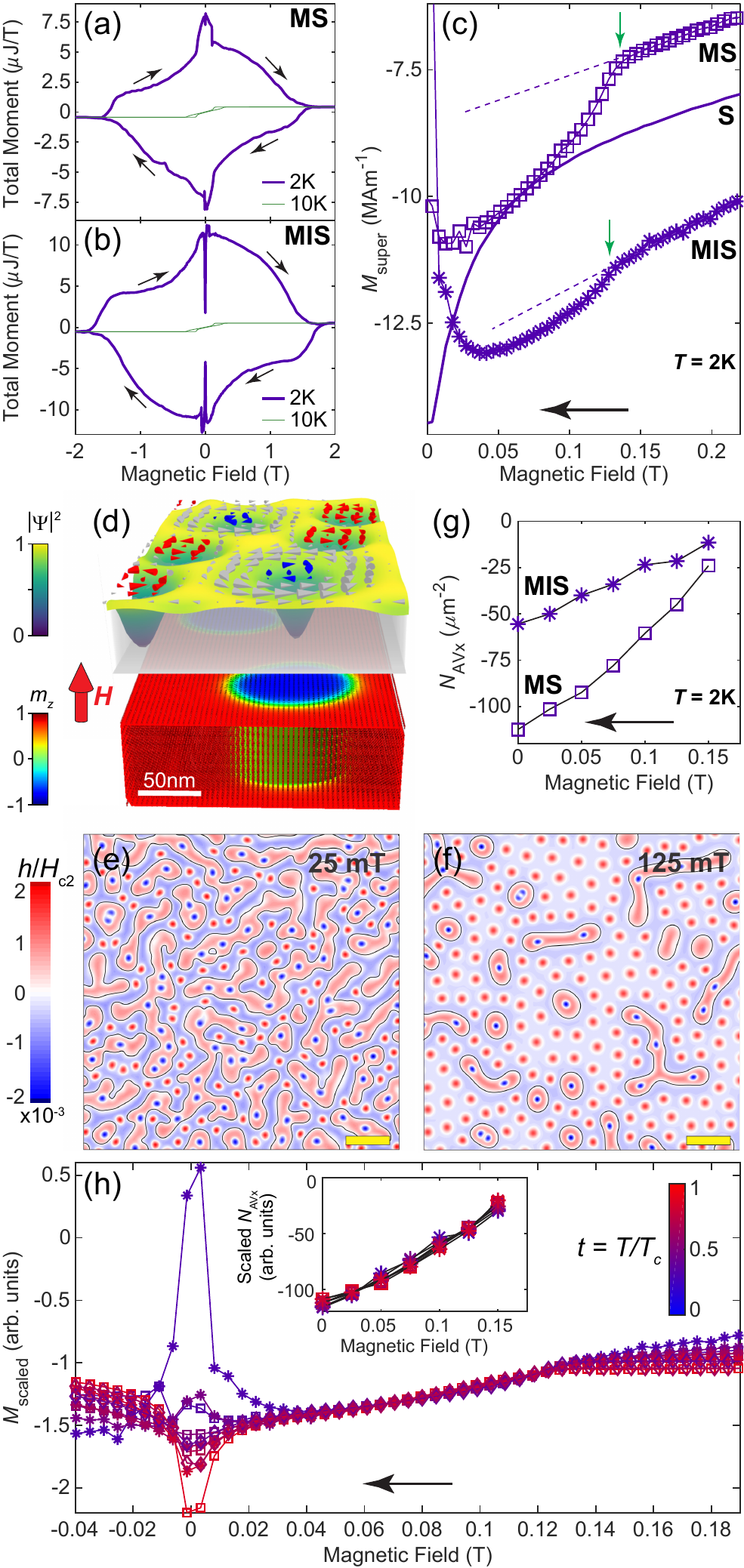}
\end{figure}

\begin{figure}[b]
\caption{\label{Fig2} Magnetization loops at 2\,K, 10\,K for (a) a MS sample ($T_c=5.75$\,K) and (b) a MIS sample ($T_c=6.25$\,K). Black arrows indicate field sweep directions. (c) Evolution of the superconducting magnetization $M_{\mathrm{super}}(H)$ through $H_\mathrm{nuc}$, extracted by subtracting $m(H,T=10\,\mathrm{K})$ from $m(H,T<T_c)$. Data from a reference 25\,nm Nb film are labeled S. Green arrows and dashed lines highlight the change in $M_{\mathrm{super}}(H)$ below $H_\mathrm{nuc}$. For complete $T$-dependent $M_{\mathrm{super}}(H)$ loops, see SM V. (d) Calculated supercurrent profile above two N{\'e}el skyrmions at $H=$\,125 mT. Blue arrows indicate antivortex currents induced above skyrmion cores; red arrows depict vortex currents outside skyrmion domains. Vortex-antivortex annihilation is prevented by supercurrents screening the skyrmion stray field (grey arrows), maximal at the skyrmion domain wall. (e,f) Supercurrent-induced field $h$ simulated at $H=$~25,125\,mT, where vortices/antivortices are visible as red/blue dots and solid lines show the contours of magnetic domain walls. Scale bars are 250\,nm; simulations from $150\,\rightarrow\,-200$\,mT are shown in SM VI\,C. (g) Field-dependent simulated antivortex density $N_\mathrm{AVx}$ for MS and MIS heterostructures. (h) Scaled magnetization $M_{\mathrm{scaled}}=aM_{\mathrm{super}}+b$ in MS samples ($\square$,$\lozenge$) and a MIS sample ($\ast$) for 2\,K\,$<T<$\,5\,K. The inset shows a similar scaling for $N_{\mathrm{AVx}}(H)$.
}
\end{figure}

\textit{Superconducting layer.}\textemdash~Knowing $r_{sk}$, we can tune the Nb layer thickness to optimize skyrmion-vortex coupling. In films of thickness $d_s < \lambda$, Meissner screening is weak and $\lambda$ is replaced by the Pearl depth $\Lambda = \lambda^2/d_s$~\cite{Pearl1964}. To ensure long-range vortex interactions and negligible Nb bulk pinning, we select $d_s$ = 25\,nm, resulting in $\Lambda\gtrsim200$\,nm (SM III). Figure~\ref{Fig1}(j-l) display the low-field vortex matter in a 25\,nm Nb film with $T_c=6.05$\,K, imaged by scanning tunneling spectroscopy (STS). The spatial evolution of the zero-bias conductance suggests a coherence length $\xi_{eff}(0.4\mathrm{K})\approx36$\,nm (SM IV), far exceeding the Ginzburg-Landau (GL) $\xi(0)\approx10$\,nm (SM III). This disparity likely originates from vortex core expansion in the Pt layer encapsulating the Nb against oxidation~\cite{Stolyarov2018}. Our heterostructures thus satisfy $\xi<\xi_{eff}<r_{sk}<\Lambda$.

\textit{Antivortices induced by skyrmions.}\textemdash~Figure~\ref{Fig2}(a) shows the total moment $m(H)$ of a MS heterostructure.  Above $T_c$, $m(H)$ displays the expected hysteresis from the magnetic layer; below $T_c$ superconductivity dominates. A MIS heterostructure [Fig.~\ref{Fig2}(b)] exhibits similar behavior. In Fig.~\ref{Fig2}(c), we isolate the superconducting component $M_{\mathrm{super}}(H)$ and track its evolution while decreasing $H$. Below $H_\mathrm{nuc}$, the chiral film minimizes its free energy by nucleating (negatively magnetized) skyrmionic domains. This coincides with $M_{\mathrm{super}}(H)$ turning sharply towards increasingly negative values in both MS and MIS samples. In contrast, $M_{\mathrm{super}}(H)$ of a bare 25\,nm Nb film (S) evolves smoothly through this field range. The change in $M_{\mathrm{super}}$ below $H_\mathrm{nuc}$ indicates two processes in our heterostructures: (1) ejection of vortices (with moment $m>0$) and (2) antivortex formation with $m<0$ (parallel to the skyrmion cores).

\begin{figure}[t]
\includegraphics[clip=true, width=\columnwidth]{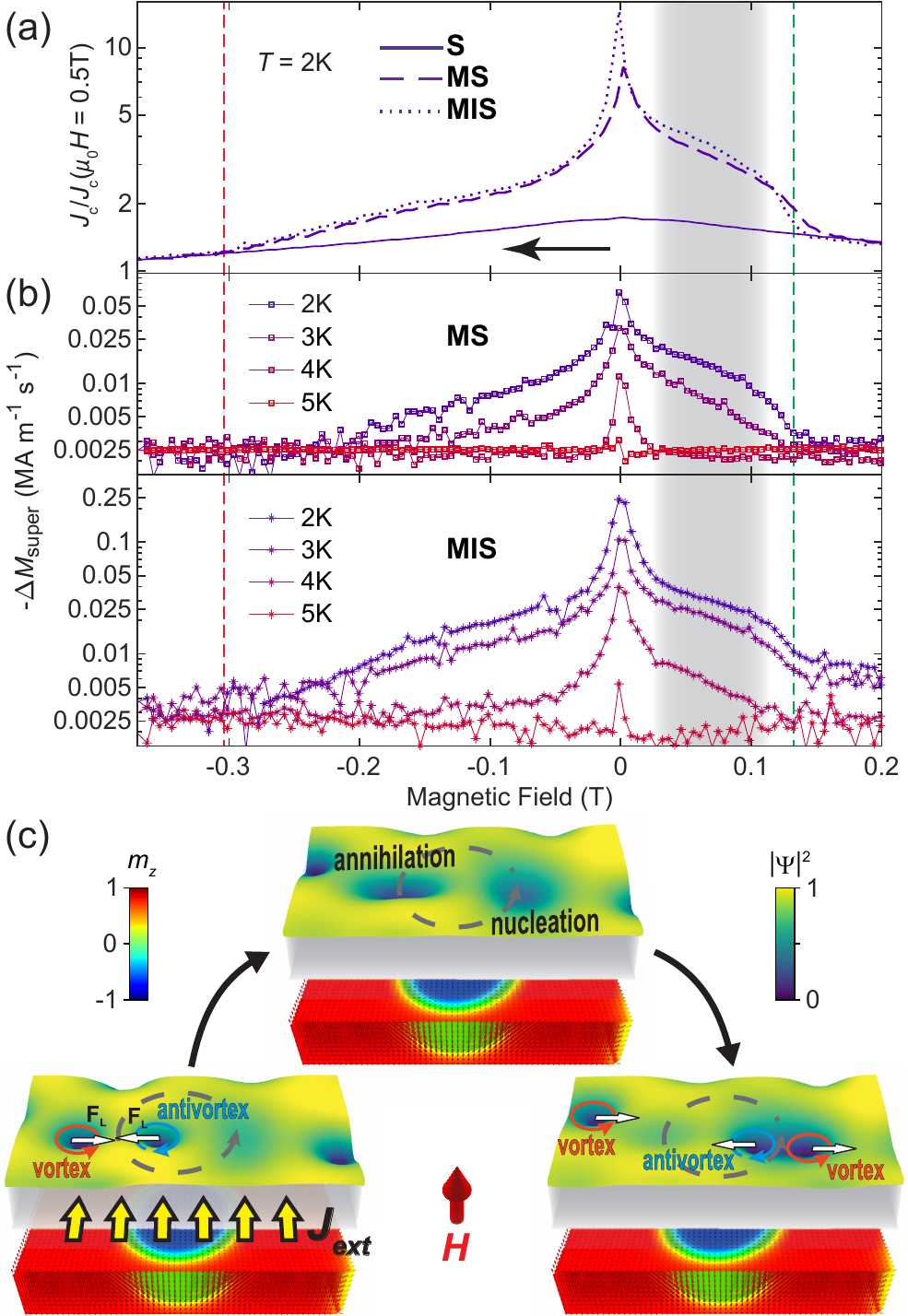}
\caption{\label{Fig3}(a) Field-dependent critical currents $J_c(H)$ in MS, MIS heterostructures and a 25\,nm Nb film (S), extracted from $V(I)$ curves at $T=2$\,K (SM VII). The arrow shows the field sweep direction. $H_\mathrm{nuc},H_\mathrm{ann}$ from Fig.~\ref{Fig1}(c) are highlighted by green and red dashed lines. (b) Thermal evolution of the magnetic relaxation $\Delta M_{\mathrm{super}}(H)$ in MS and MIS heterostructures. $\Delta M_{\mathrm{super}}(H)$ is the change in $M_{\mathrm{super}}$ in a 60\,s period after changing the field. (c) Antivortex-facilitated vortex tunneling through chiral domains. The Lorentz force ${F_L}$ from the applied current $J_{ext}$ acts in opposite directions for vortices/antivortices.  Consequently, vortex-antivortex pairs mutually annihilate on one side of the skyrmion/stripe, before spontaneously re-emerging on the opposite side.  
}
\end{figure}

The saturation magnetization of our chiral film $\mu_0M_s=1.82$\,T exceeds the lower critical field of the Nb film $H_{c1} \approx 0.012$\,T at 2\,K (SM III,~\cite{Pinto2018}), indicating that magnetic textures may form (anti)vortices in the superconductor~\cite{Baumard2019}. Spontaneous vortex formation is well-established in superconductor-\emph{ferromagnet} hybrids~\cite{Lyuksyutov2005,Aladyshkin2009,Iavarone2011,Bobba2014}, but far more challenging to achieve using chiral magnets since $r_{sk}$ can be orders of magnitude smaller than typical ferromagnetic domains. Modelling the radial spin evolution as $m_z\sim\mathrm{tanh}\frac{\pi(r-r_{sk})}{r_{sk}}$~\cite{Heide2008}, we estimate an upper limit of -4.44\,$\Phi_0$ for the flux through our 50\,nm N{\'e}el skyrmions. Although the skyrmion stray field decays rapidly outside the magnetic layer, the flux piercing the adjacent superconductor remains sufficient to create antivortices in all our heterostructures (SM VI\,A). A threshold $M_s\geq\Phi_0\mathrm{ln} (\Lambda/\xi)/(0.86\pi^2d_mr_{sk})$ was derived in ref.~\onlinecite{Dahir2019} for vortex formation by zero-field skyrmions, which our heterostructures exceed by 34\% at 2\,K. As a definite proof of antivortex nucleation in our experiments, we combine micromagnetic simulations of the [Ir$_{1}$Fe$_{0.5}$Co$_{0.5}$Pt$_1$]$^{10}$ stray field (SM VI\,A) with GL simulations of the superfluid ($\left|\Psi\right|^2$) and supercurrent densities in an adjacent 25\,nm Nb film (SM VI\,B,C). Figure~\ref{Fig2}(d) shows the supercurrent and $\left|\Psi\right|^2$ profiles above two N{\'e}el skyrmions stabilized at $H =$~125\,mT, confirming the coexistence of skyrmion-induced antivortices (blue arrows) with vortices parallel to $H$ (red arrows), separated by screening currents (grey arrows) above the skyrmion domain walls. In Fig.~\ref{Fig2}(e,f) we present large simulations of vortex-antivortex states at $H=25,125$\,mT, visualized using the supercurrent-induced field $h$. At all fields, skyrmions and stripe domains [solid lines in Fig.~\ref{Fig2}(e,f)] robustly generate antivortices. 

As $H$ falls, the simulated antivortex density $N_\mathrm{AVx}$ rises [Fig.~\ref{Fig2}(g)]. MS heterostructures develop a higher $N_\mathrm{AVx}$ than MIS samples, since they experience a stronger stray field (as the Nb layer is 7~nm closer to the magnetic film). This higher $N_\mathrm{AVx}$ is visible in experiments [Fig.~\ref{Fig2}(c)] as a larger drop in $M_\mathrm{super}$ below $H_\mathrm{nuc}$ for the MS sample. Raising the temperature also increases $N_\mathrm{AVx}$, since $H_{c1}$ falls. For $T<T_c$, $M_{\mathrm{super}}(H)$ in all our heterostructures collapse onto a single curve between 0.04-0.14\,T following a linear $aM_{\mathrm{super}}+b$ scaling [Fig.~\ref{Fig2}(h)]. $N_\mathrm{AVx}(H,T)$ can be similarly rescaled onto a single curve (Fig.~\ref{Fig2}(h) inset), confirming that antivortex nucleation driven by evolving spin topology is responsible for the magnetic response of the superconductor. 

\textit{Skyrmion impact on vortex dynamics.}\textemdash~The critical current density $J_c(H)$ measured by electrical transport (SM VII) is plotted in Fig.~\ref{Fig3}(a). MS and MIS heterostructures both exhibit an enhanced $J_c(H_\mathrm{nuc}{\rightarrow}H_\mathrm{ann})$ relative to a bare Nb film. This behaviour is mirrored by an emergent time-dependent magnetization [Fig.~\ref{Fig3}(b)]. $M_{\mathrm{super}}(H_\mathrm{nuc}{\rightarrow}H_\mathrm{ann})$ relaxes towards increasingly negative magnetizations over several minutes after every reduction in $H$, leading to a measurable drop $\Delta M_{\mathrm{super}}$. This relaxation and the rise in $J_c$ are caused by the chiral magnet inducing supercurrents which impede vortex motion. Just below $H_\mathrm{nuc}$, the skyrmion/antivortex density is low and the initial rise in $J_c(H)$ originates from long-range vortex-antivortex attraction. Reducing the field increases the skyrmion density, obliging vortices moving in a flux gradient or applied current to cross chiral domains. Vortices are repelled by skyrmion screening currents [Fig.~\ref{Fig2}(d)]; however the presence of an antivortex reduces the barrier height, enabling vortices to cross the skyrmion/stripe in a process analogous to Klein tunneling. Flux-flow simulations are shown in Fig.~\ref{Fig3}(c), SM VI\,D and Supplemental videos. Since $N_\mathrm{AVx}$ is larger in MS heterostructures, the vortex mobility is enhanced, reducing their $J_c$ and $\Delta M_{\mathrm{super}}$ relative to MIS samples at intermediate fields [grey shading in Fig.~\ref{Fig3}(a,b)]. Around zero field, a sharp maximum in $J_c$ and $\Delta M_{\mathrm{super}}$ originates from strong vortex pinning by labyrinthine stripes (SM II\,C,~\cite{Palermo2020}). This phenomenon also causes the upturn and scaling failure in $M_{\mathrm{super}}(H)$ below 0.04\,T [Fig.~\ref{Fig2}(c,h)]. However, the $J_c,\Delta M_{\mathrm{super}}$ maxima in MS heterostructures are $>50$\% lower than in MIS samples, confirming the crucial role of antivortices in facilitating vortex motion.

\begin{figure}[tbp]
\includegraphics[clip=true, width=\columnwidth]{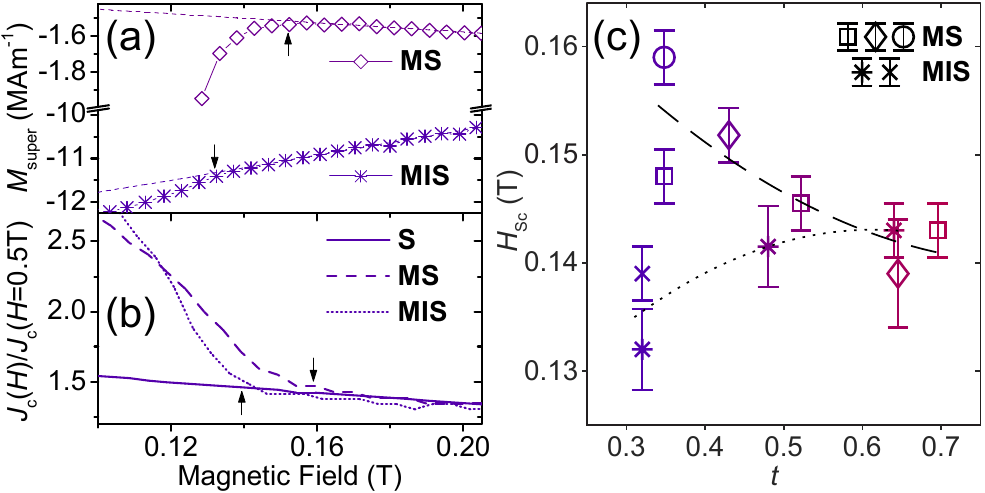}
\caption{\label{Fig4}(a) $M_{\mathrm{super}}(H)$ comparison between MS and MIS samples for a downward field sweep through $H_\mathrm{nuc}$ at $T=2$~K. Dashed lines are linear fits in the 0.17~T~$<H<$~0.22~T range; arrows indicate the response field $H_\mathrm{Sc}$ [defined by a 1\% deviation from linearity in $M_{\mathrm{super}}(H)$]. (b) Comparison of normalized $J_c(H)$ in MS and MIS samples during a field sweep at $T=2$~K.  Arrows indicate $J_c(H)$ rising above the background (defined by a S film) at $H_\mathrm{Sc}$. (c) Thermal evolution of $H_\mathrm{Sc}$ probed by magnetization ($\square$,$\lozenge$,$\ast$) and transport ($\circ$,$\times$); see SM VIII for source data and details. Dashed/dotted lines highlight the trends displayed by MS/MIS bilayers.  
}
\end{figure}

\textit{Skyrmion-vortex exchange coupling.}\textemdash~Our STS data (Fig.~\ref{Fig1}(j-l), SM IV) suggest $\xi_{eff}(0.4\mathrm{K})\approx36$\,nm in the 2\,nm Pt separating Nb from [Ir$_1$Fe$_{0.5}$Co$_{0.5}$Pt$_1$]$^{10}$. This corresponds to $\xi_{eff}(2\,\mathrm{K})\approx43$\,nm, close to $r_{sk}\approx50$\,nm. Exchange coupling will consequently be weak, since the induced Rashba-Edelstein currents centred at $r=r_{sk}$ scarcely interact with antivortex currents which peak at $r\approx2\xi_{eff}$. However, MFM images close to $H_\mathrm{nuc}$ show larger skyrmions with $r_{sk}$ up to $\sim60$~nm (Fig.\ref{Fig1}(e), SM II\,B). Since any increase in $r_{sk}$ should enhance exchange coupling, we search for differences between MS and MIS heterostructures at high fields. Figure~\ref{Fig4}(a) shows that the field at which $M_{\mathrm{super}}$ first responds to skyrmion formation (denoted $H_\mathrm{Sc}$) is $\sim$\,20\,mT higher in MS versus MIS samples at 2\,K. This pattern is repeated in the critical current [Fig.~\ref{Fig4}(b)], where MS samples exhibit an earlier and larger $J_c(H)$ enhancement in the 0.16 $\rightarrow$ 0.12~T field range. Since the flux lattice stray field has negligible influence on skyrmion formation (SM IX), we interpret these data as exchange coupling assisting antivortex creation at high fields, by inducing supercurrents rotating in the same sense as antivortex currents. As the temperature rises the difference in $H_\mathrm{Sc}$ between MS and MIS heterostructures is suppressed [Fig.~\ref{Fig4}(c)], due to the divergence in $\xi$ which only enables the $r_{sk}>\xi_{eff}$ condition to be fulfilled at low temperature. Exchange coupling can be strengthened in future hybrids by minimizing $\xi_{eff}$ at the superconductor/chiral magnet interface.  

\textit{Summary.}\textemdash~We have shown that skyrmion stray fields can nucleate stable antivortices in engineered chiral magnet-superconductor heterostructures. This process is independent of SOC, thus simplifying the task of building a topological superconductivity platform from (anti)vortices coupled to chiral spin textures. Coexistence of these hybrid magnetic solitons with superconducting vortices creates a complex yet controllable environment for exploring unique emergent flux dynamics. 

\section{Acknowledgements}
We acknowledge support from the Singapore National Research Foundation (NRF) NRF-Investigatorship (No. NRFNRFI2015-04) and Singapore MOE Academic Research Fund Tier 3 Grant MOE2018-T3-1-002. M.\,Raju thanks the National University of Singapore E6NanoFab laboratory and the Data Storage Institute, Singapore, for access to sample growth facilities. C.R. acknowledges financial support from the Swiss National Science Foundation (Grant No.\,182652) and thanks A. Guipet for technical assistance. The work of R.M.M., M.J.W. and M.V.M. was supported by Research Foundation Flanders (FWO), the University of Antwerp (BOF), and the VSC (Flemish Supercomputer Center), funded by the FWO and the Flemish Government - department EWI. The collaboration in this work has been fostered in part by EU-COST Action CA16218 NANOCOHYBRI.

%

\end{document}